\documentclass[aip,pof,reprint,onecolumn]{revtex4-1}
\usepackage{graphicx,epstopdf}

\begin{document}
\title{Transition to turbulence in slowly divergent pipe flow}
\author{Jorge Peixinho$^{1,2}$ and Hugues Besnard}
\affiliation{$^1$Laboratoire Ondes et Milieux Complexes, CNRS and Universit\'e du Havre, 53 rue de Prony, 76600 Le Havre, France \\
$^2$Fluid Engineering Laboratory, Department of Mechanical Engineering, University of Tokyo, 7-3-1 Hongo, Tokyo, 113-8656, Japan}
\date{\today}

\begin{abstract}
The results of a combined experimental and numerical study of the flow in slowly diverging pipes are presented. Interestingly, an axisymmetric conical recirculation cell has been observed. The conditions for its existence and the length of the cell are simulated for a range of diverging angles. There is a critical velocity for the appearance of this state. When the flow rate increases further, a subcritical transition for localized turbulence arises. The transition and relaminarization experiments described here quantify the extent of turbulence. The findings suggest that the transition scenario in slowly diverging pipes is a combination of stages similar to those observed in sudden expansions and in straight circular pipe flow.
\end{abstract}
\maketitle

\begin{figure}
\includegraphics[width=0.35\columnwidth]{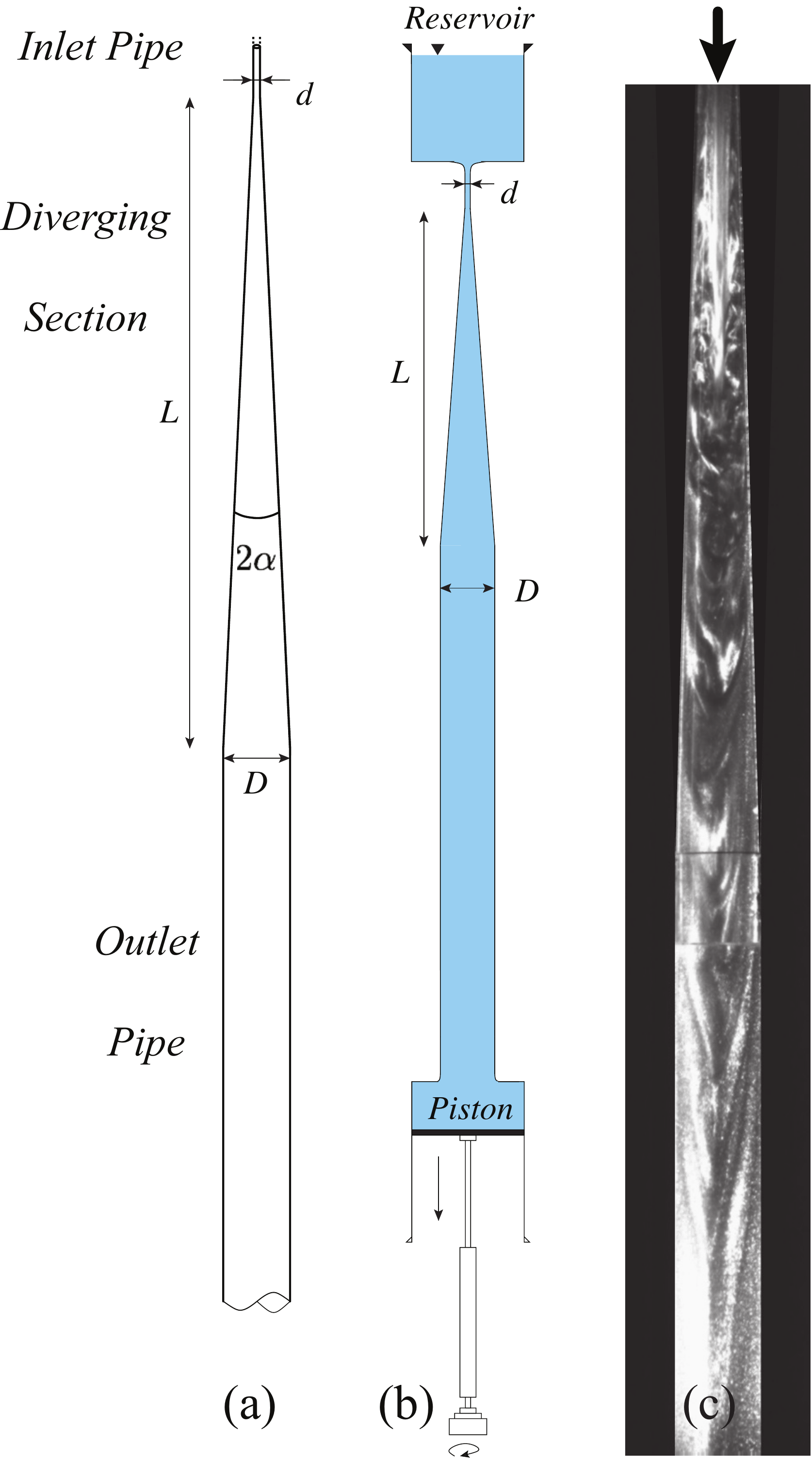}
\caption{Schematics of the diverging pipe, the experiment and a flow visualization photograph. (a) Basic geometry, (b) sketch of the experimental setup drawn up to scale, and (c) flow visualization photograph for $Re=1000$ in the diverging pipe ($\alpha=2^\circ$, $\beta=L/d=128.88$, and $D/d=8.79$). The flow is from top to bottom and the horizontal lines on the photograph indicate the connection between the divergent section and the expansion (Enhanced Online).}
\label{fig1}
\end{figure}

The flow in slowly diverging pipes, {\it{i.e.}} cylindrical pipes of slowly increasing diameter along the pipe axis, as depicted in figure \ref{fig1}(a), are not well documented despite some fundamental and practical features. This flow arises in microfluidics when transferring liquid using pipettes and in physiological flows in veins when blood rushes from organs and tissues towards the heart. Knowledge of the nature of this flow can also be useful in the context of burner-combustion systems, jet engine exhaust, thrust-vectoring nozzles, and flows of confined jets.\cite{craya55,soong98,nahum06}

The general two-dimensional problem of flow stability between two plane walls meeting at a source point with an angle is known as the Jeffery-Hamel problem. There are several theoretical and numerical developments where bifurcations have been found \cite{sobey86,hamadiche94,kerswell04,stow01,haines11,swaminathan11} and all these works indicate a rich and diverse set of solutions even for small diverging angles. Recently, Putkaradze and Vorobieff \cite{putkaradze06} observed, using particle image velocimetry, the multiple vortex flow r\'egime predicted by Kerswell {\it et al.} \cite{kerswell04}

A large body of research has dealt with two-dimensional sudden channel expansions, with sharp 90$^\circ$ corners. In the case of a 1:3 sudden expansion flow, it was shown \cite{durst74,fearn90} that the asymmetry arises at a critical Reynolds number through a pitchfork symmetry breaking bifurcation. Fearn {\it et al.} \cite{fearn90} were able to measure the degree of asymmetry due to small imperfections of the experimental apparatus and compared it with numerical results. 

The present work considers the case of an axisymmetric circular pipe that is slowly expanding. Solutions for the laminar flow in slightly tapered cylinders assuming the lubrication approximation $\left(D-d\ll L\right)$ can be found in Bird {\it et al.} \cite{bird87} in the form of corrected expressions for the velocity profiles. Here $d$ and $D$ are the inlet and outlet diameter of the divergent section and $L$ is its length (see figure \ref{fig1}(a)). The diverging angle, $\alpha$, refers to the half angle of the diffuser. When the diameter varies slowly, the axial velocity profile, which depends on the local diameter, conserves its parabolic shape. However, in the diverging section, the centerline velocity scales as $1/x^2$, where $x$ is the axial position, and a fluid particle experiences a rapid deceleration. Additionally, in laminar flow, the pressure along the diverging section decreases rapidly.\cite{rosa06}

The transition to turbulence in the limit case of a $90^\circ$ (abrupt) 1:2 circular pipe expansion was studied by Sreenivasan and Strykowski \cite{sreenivasan83}, Latornell and Pollard \cite{latornell86} and others. The steady flow becomes unstable and a periodic time-dependent state was observed at $Re\sim 750$ and around $1500$.  $Re$ is the Reynolds number based on $d$. In a modern investigation using high resolution magnetic resonance imaging, Mullin {\it et al.} \cite{mullin09} found a sharp onset of asymmetry in the downstream flow at $Re=1140$. Recently, numerical simulations \cite{sanmiguel-rojas10,cantwell10} confirmed the previous experimental results and showed that the flow is unstable to infinitesimal perturbation for $Re=3273$. The exact nature of the first instability is unclear since imperfections are likely to produce disturbance that can grow and lead to multiple solutions both stationary and time-dependent.

The stability of the axisymmetric slowly diverging pipe flow has been investigated in a numerical work \cite{sahu05} solving a multigrid Poisson equation for the base flow and partial differential equations for the flow stability in a diverging pipe ($\alpha=1.5^\circ$, $L/d=120$ and $D/d\approx7.3$) and indicate the flow  is linearly unstable from $Re=150$.  In another numerical simulation \cite{sparrow09} it is found that the flow separation occurs for $Re$ less than about 2000. The purpose of the present experimental study is to add new quantitative data in order to clarify these findings. 

A number of investigations in straight pipes with a axisymmetric constriction have been carried out numerically\cite{sherwin05} and in experiments \cite{vetel08,griffith08} as an idealization of a stenosed artery. In this situation, the velocity profile at the inlet of the divergent section is almost flat, the flow in this divergent pipe ($\alpha\approx45^\circ$, $L\approx D$ and $D/d=2$) exhibits a laminar recirculation region and the subcritical transition to turbulence\cite{sherwin05} occurs at $Re=361$.

\begin{figure}[h]
\includegraphics[width=1.0\columnwidth]{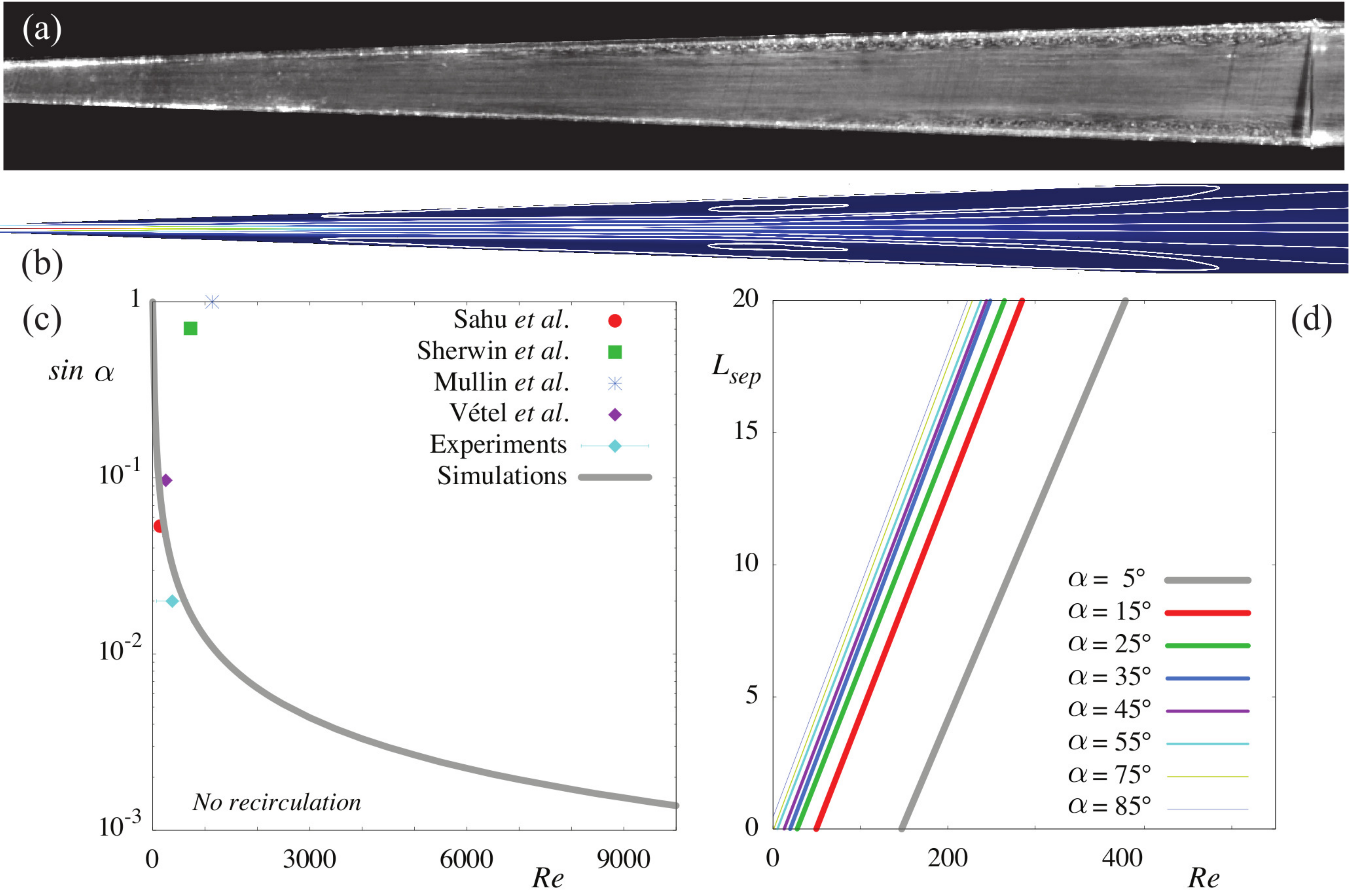}
\caption{Laminar flow in a slowly divergent pipe. (a) Time-exposure photograph for $Re = 200$. (b) Numerical simulations of velocity streamlines for $Re=600$. (c) Onset of axisymmetric recirculation cell as predicted from numerical simulations in terms of $\sin \alpha$  as a function of $Re$. The points represent the critical $Re$ of transition to turbulence for the current and previous studies\cite{mullin09,sherwin05,vetel08,sahu05}. (d) Numerical predictions of the separation length, $L_{sep}$, versus $Re$ for different $\alpha$}
\label{fig2}
\end{figure}

The following section of this paper presents the experimental apparatus. In \S 3 the results of two-dimensional simulations, as well as an investigation of the stability of the recirculation cells are given. In \S 4 the dynamics of the turbulent patches is described and these are tested via a series of relaminarization experiments. Conclusions are drawn in \S 5. 

The experiments consist of flow visualization in slowly diverging axisymmetric pipes. A schematic of the experimental setup is given in figure \ref{fig1}(b). It is composed of a vertical pipe made of acrylic. The flow is controlled using a syringe pump (TSE Systems Model 540230) together with 100 ml glass syringes. The device pulls the fluid at a constant mass flux along the pipe. Hence, even if the motion becomes turbulent, the mass flux through the pipe is unaffected so that $Re$ remains constant. The maximum pulling velocity corresponded to $Re=4000$.

The slowly diverging acrylic pipe used here has an angle, $\alpha$, of $2\,^\circ$ (or $\pi/90$ radians) over a length $L=128.88d=275.8$ mm. The inlet diameter was $d=2.14\pm0.1$ mm, the outlet diameter was $D=18.8\pm0.1$ mm and there was an imperfection due to the fitting of the outlet section. Downstream of the divergent section, a straight pipe section extended for $320d$.

The inlet has a straight section of diameter $d$ over $10d$ in order to obtain a fully developed Poiseuille flow. The development is facilitated by a smooth contraction between the inlet and the reservoir. For flow visualization, 4 mL of Kalliroscope, a suspension of reflective flakes, was added to 2 liters of degassed water. A vertical light sheet was formed in the center plane of the flow, and a camera was used to record the dynamics of the flow. An example of the flow visualization is shown in figure \ref{fig1}(c). The Reynolds number is defined by: $Re=Ud/\nu$ where $U$ is the mean flow rate and $\nu$ is the kinematic viscosity. The temperature of the fluid is taken into account in the calculation of $Re$. The other parameters of the diverging pipe are the expansion ratio: $E=D/d$ between the outlet and inlet diameters and the non-dimensional length of the diverging section:  $\beta=L/d$. For the $2^\circ$ pipe, $\left(E,\beta\right)=\left(8.79,128.88\right)$.

The base flow in slowly diverging pipes is a parabolic velocity profile and the axial velocity on the centerline decreases along the divergent axes. As the flow rate increases or as the diverging angle increases, a recirculation cell is observed close to the walls and its extent depends on $Re$. In practice, the cell is thin and difficult to measure because of the azimuthal curvature of the outer walls. An example of time-exposure photography is presented in figure \ref{fig2}($a$). The fluid particle paths around the centerline appear as continuous lines whereas fluid particle paths close to the walls appear as dotted lines indicating that they move at a much slower pace. 

The axisymmetric flow is reproduced in numerical simulations of time-dependent Navier-Stokes equations using a two dimensional axisymmetric finite element code (COMSOL Multiphysics). At the inlet, the velocity profile is parabolic over $10d$. In the downstream section, the mesh consists of several blocks and is sufficiently long, typically $100d$, so that Poiseuille flow is recovered. The outlet boundary condition is constant pressure. The number of elements along the divergent section and the outlet section depends on $\alpha$ and $E$ and is around one million. The numerical simulations indicate that the recirculation appears at a finite $Re$ in the corner between the divergent section and the outlet section. As the flow rate increases, the recirculation cell grows both upstream and downstream. The calculated streamlines as well as axial velocity magnitude are given in figure \ref{fig2}(b). The flow can be described as a confined jet which expands downstream and the recirculation cell is long and thin. Notice that the velocity profile exhibits a flow reversal containing inflection points which can lead to Kelvin-Helmholtz instability.

The onset for the growth of the recirculation cells can be tracked using numerical simulations for a range of $\alpha$, $\beta$, and $E$. For $\alpha\approx 40$ up to $180^\circ$ ($\sin \alpha \approx 0.34$ up to 1), the recirculation cell is always present. As $\alpha$ decreases, there is a critical $Re$ for the onset of the recirculation cell which grows rapidly as shown in figure \ref{fig2}(c). Eventually, when $\alpha$ tends to zero, the critical $Re$ becomes large as suggested by linear stability calculations of circular pipe flow of constant diameter \cite{meseguer03}. The threshold for the appearance of the recirculation cell is compared to data points for transition to turbulence from the literature.\cite{mullin09,sherwin05,vetel08,sahu05} Calculations were performed changing $\alpha$ (or $\beta$) keeping $E$ constant from 2 to 10 and the critical $Re$ for the onset of recirculation cell varies by less than 5\%.

In the regions around the outlet of the divergent section, the pressure close to the wall and the deceleration of the fluid particle lead to the onset of the recirculation cell. Once the cell appears, it grows linearly with $Re$. In figure \ref{fig2}(d), the axial length of the recirculation bubble, $L_{sep}$, called  the separation length, deduced from the positions of zero wall shear stress calculations is presented as a function of $Re$ for different $\alpha$. Our numerical simulation for $\alpha=85^o$ agree quantitatively with the experiments of Latornell and Pollard\cite{latornell86} on sudden expansion $\left(L_{sep}=0.096Re\right)$ and previous numerical works.\cite{cantwell10,griffith08}

\begin{figure}[t]
\includegraphics[width=0.98\columnwidth]{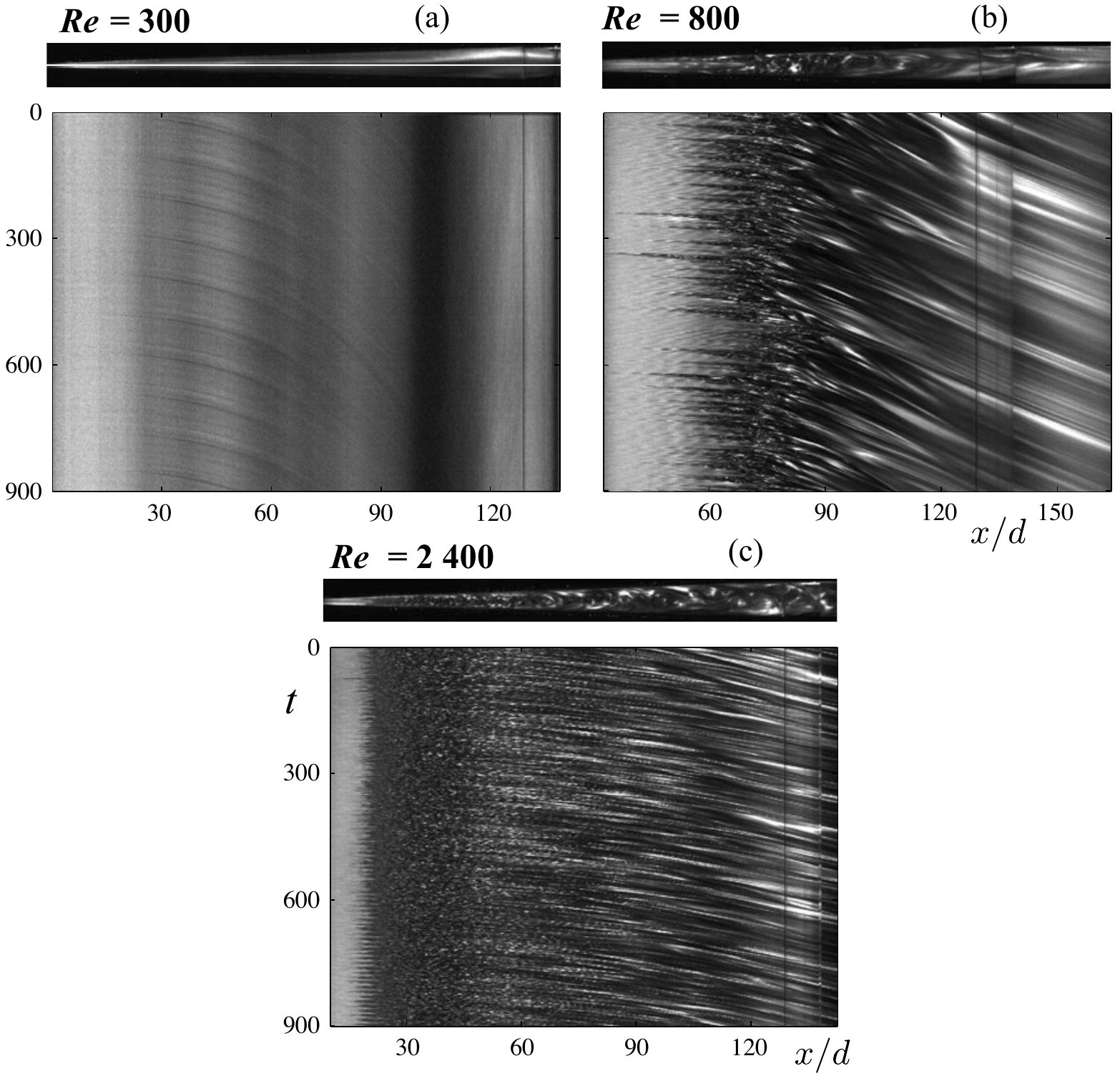}
\caption{Photographs of the flow and space-time diagrams for different $Re$. The photographs are on top of the diagrams, which are based on the brightness along the flow axis.  (a) Laminar flow at $Re=300$, (b) localized turbulent patch at $Re=800$, and (c) turbulent patch at $Re=2400$}
\label{fig3}
\end{figure}

In the case of abrupt expansions in two-dimensional channels, several authors refer to the ``Coanda'' effect\cite{fearn90,soong98} when the initial symmetric flow becomes asymmetric. Specifically, one of the recirculating cells becomes larger and a new time-independent flow is observed. This sequence of events leads to the breaking of the symmetry of the flow in agreement with the ideas of bifurcation theory. In the present case of axisymmetric flow there is a single recirculation cell in contrast with the two-dimensional channels where there is one cell behind each step. Sanmiguel-Rojas and Mullin\cite{sanmiguel-rojas10} showed using three-dimensional numerical simulations that the axisymmetric state is sensitive to small imperfections. Depending on the amplitude of a small distortion added to the parabolic inlet flow, the flow change to asymmetric or disordered time-dependent state. Our numerical simulations are two-dimensional axisymmetric. In the experiments,  the time-independent asymmetric states were not clearly observed because of azimuthal curvature of the outer walls (see figure \ref{fig2}(a)). At higher $Re$, the cell is found to be sensitive to natural disturbances of the system. These imperfections are related to a distortion parameter in a complicated manner and are the source instabilities which lead to the formation of super-critical turbulent patches.

With a further increase of the flow rate, the recirculating bubble breaks down into localized turbulent patches as the one depicted in figure \ref{fig1}(c). These localized turbulent patches have some similarities with the so-called localized puffs observed in cylindrical straight pipe flow\cite{wygnanski73}. Puffs seem to have a definite length for a given $Re$, an active core of high turbulence intensity and a decaying wave at the front. In the diverging pipe outlet section, turbulent patches appear in the divergent section and extend over many diameters. There, the downstream Reynolds number, based on the outlet diameter, is too small to sustain turbulence and decaying turbulence is observed. Contrary to puffs, the turbulent patch does not travel along the pipe. Their origin is the breakdown of the recirculation cell. However they do have a definite length for a given $Re$, an active core of high turbulence intensity and a decaying wave at the front. 

\begin{figure}[t]
\includegraphics[width=0.8\columnwidth]{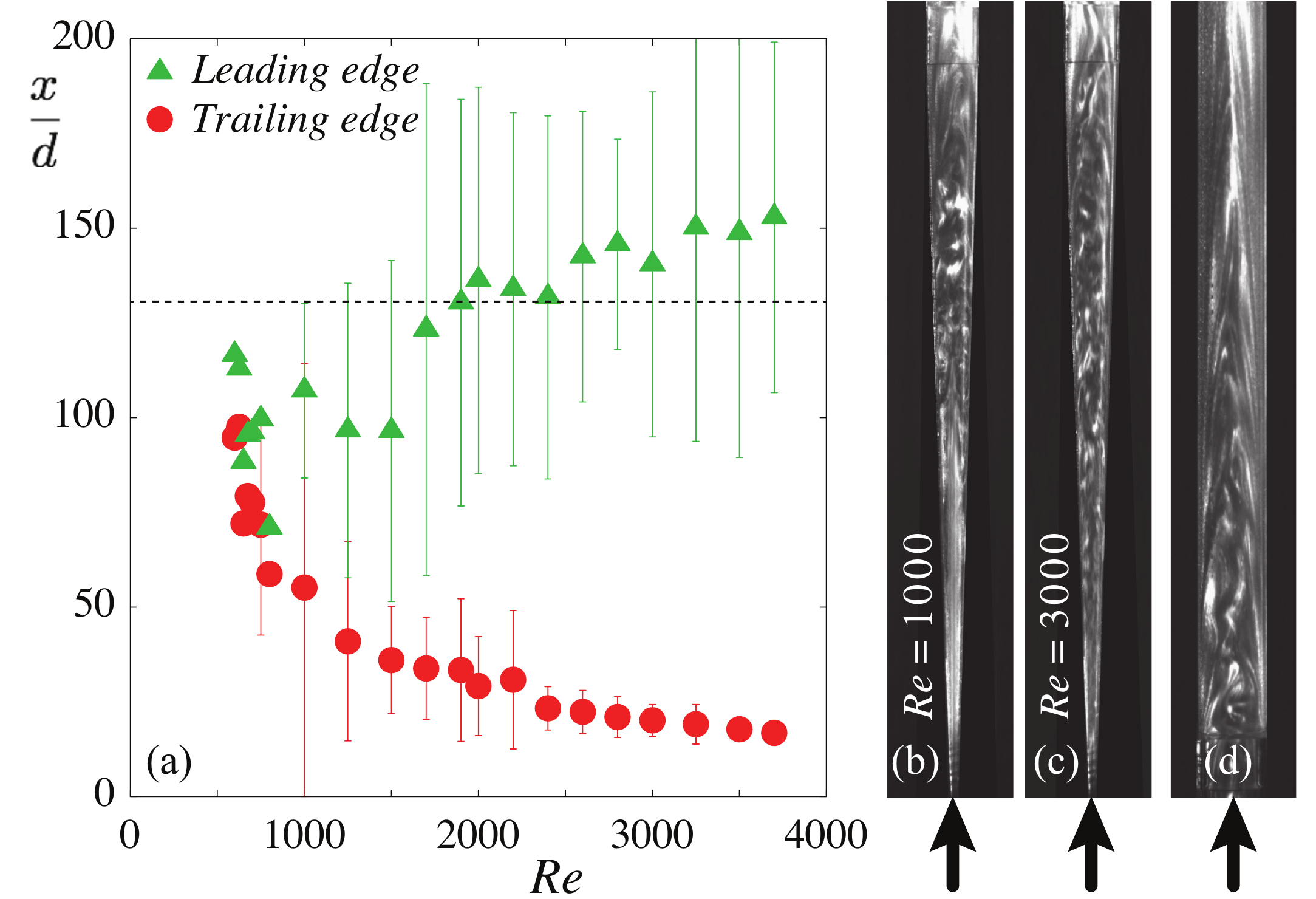}
\caption{Length of the turbulent patch in the slowly diverging pipe. (a) Position of the leading and trailing edges of the turbulent patch as a function of $Re$. The dashed line represent the end of the diverging section at $\beta=x/d=128.88$. (b) Flow visualization of a turbulent patch at $Re=1000$, (c) flow visualization of a turbulent patch at $Re=3000$ in the divergent section, and (d) its decaying front wave in the outlet section}
\label{fig4}
\end{figure}

In figure \ref{fig3}, space-time diagrams of laminar flow (figure \ref{fig3}(a)) and turbulent patches (figure \ref{fig3}(b) and (c)) are produced by converting the brightness of the flow visualization photographs along the flow axis and stacking the different lines corresponding to different times. $x/d=0$ represents the inlet of the divergent section. The dimensionless time, $t$, used here is defined as  $t=f t^*$ where $f$ is the image acquisition frequency (20 Hz) and $t^*$ is the time (in seconds). In figure \ref{fig3}(b), the puff is represented by fluctuating dark areas in the middle of the diagrams: $60<x/d<100$. The comparison of the diagrams for $Re=800$ (figure \ref{fig3}(b)) and $Re=2400$ (figure \ref{fig3}(c)) suggests that the length of the turbulent patch increases. In figure \ref{fig3}(b), the clear streaks beyond $x/d=90$ indicate that constant bright regions are moving at constant velocity. The velocity is given by the slope of the streak. The appearance of the constant brightness streaks are used to detect the decaying wave and the leading edge of the turbulent patches. More information about the relation between reflected light intensity and velocity field can be obtained from Abcha {\it et al.} \cite{abcha08}. Figure \ref{fig4}(a) presents the positions of the leading and trailing edges of the turbulent patch as a function of $Re$. The error bars represent the fluctuation of the positions of the leading and trailing edges. The trailing edge is sharper than the leading edge (see figure \ref{fig4}(b), (c) and (d)). The boundary between the decaying wave and the laminar flow is tenuous as in the case of puffs\cite{wygnanski73}. As $Re$ increases, the extend of the turbulent patch increases. The position of the trailing edge asymptotes towards a positive value as the turbulent patch cannot propagate beyond the inlet of the divergent section. For large $Re$ the position of the leading edge continues to increase. The results of figure \ref{fig4} is reminiscent of Wygnanski and Champagne\cite{wygnanski73} measurements describing the growth of turbulent puffs in uniform pipes suggesting that the turbulent patches observed here may contain solutions similar to that observed in pipe flow.

It is expected that the position of the leading edge will increase as $Re$ increases. Eventually as the turbulent patch grows a puff-slug transition\cite{wygnanski73,duguet10} is likely to take place where the stationary turbulent patch will split. This process is found to be vortex shedding via a Kelvin-Helmholtz mechanism from wall-attached shear layers.

Further increase of $Re$ means a fully developed turbulent flow field. There are many reports on turbulent flow properties in rectangular diffusers suggesting that the manipulation of the recirculation can lead to changes in the conversion of mean-flow kinetic energy to pressure\cite{schneider11,herbst07}. A recent review on turbulent flow in diffusers and direct numerical simulation of the turbulence statistics and coherent structures can be found the paper by Lee {\it et al.}\cite{lee12}

In order to quantify further the turbulent patch r\'egime, relaminarization experiments\cite{peixinho06} were performed where turbulent patches are generated and its decay is observed back to laminar as $Re$ was reduced. During the decay of the turbulent patch, a laminar liquid jet going through the turbulent patch sets in quickly and induces the recirculation flow close to the wall. Wavy patterns are observed. The liquid jet seems to fold like a viscous thread. Similar oscillations of the liquid jet in a divergent section were also observed in microfluidic experiments.\cite{cubaud06}

\begin{figure}
\includegraphics[width=0.9\columnwidth]{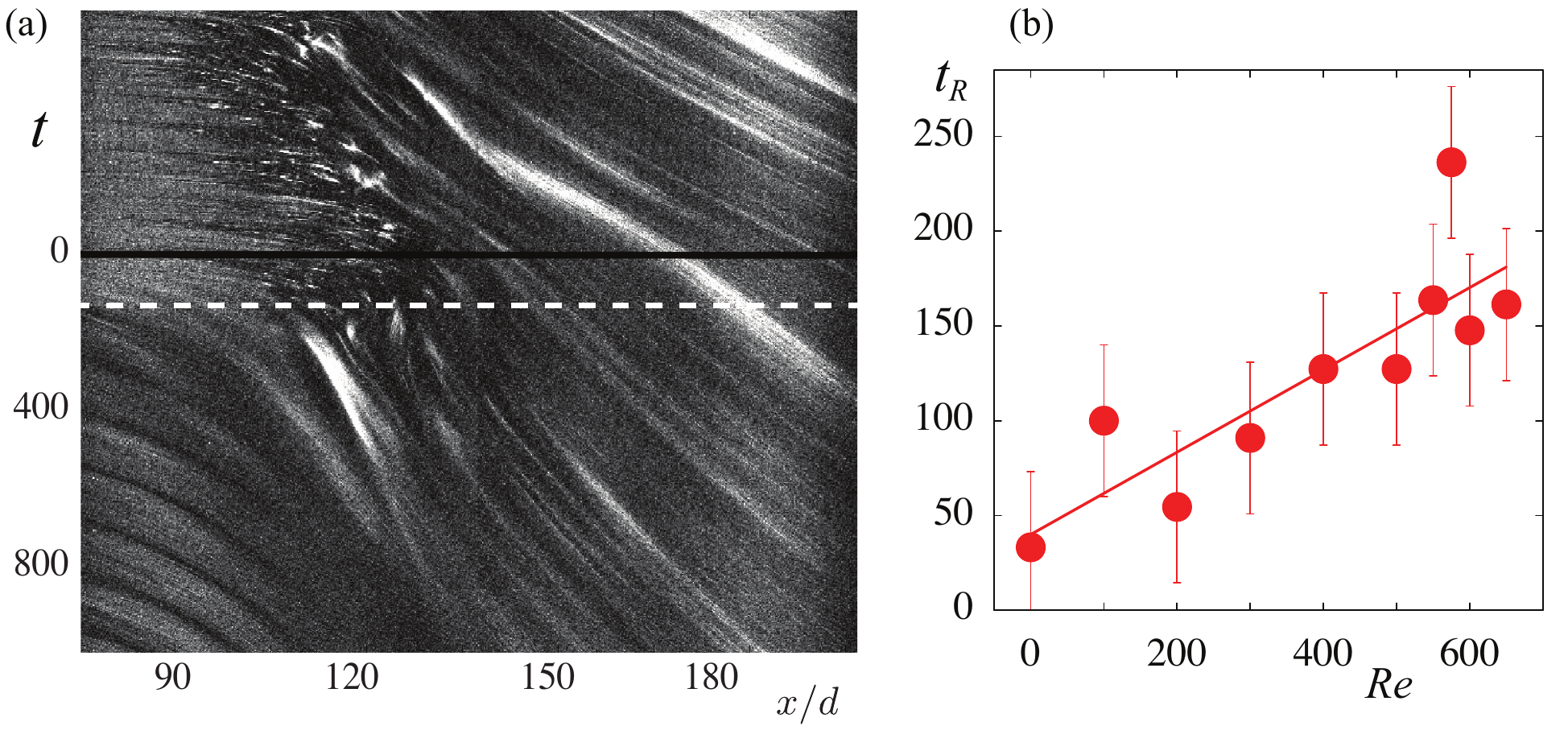}
\caption{Relaminarization experiments. (a) Space-time diagram of a relaminarization experiment from a localized turbulent patch: $Re_0=800$ to $Re=400$, the black line represents the moment of the reduction of $Re$ and the dashed white line the estimated relaminarization time when no disordered motion is observed. (b) Relaminarization time, $t_R$, versus $Re$}
\label{fig5}
\end{figure}

Our strategy was to generate a turbulent puff at $Re_0=800$. This turbulent patch was stable (see figure \ref{fig3}(b)) and is considered as a natural state or attractor of the system. The reduction in $Re$ was almost instantly and achieved by reducing the syringe pump velocity. The relaminarization was monitored through spatio-temporal diagrams. A typical example is presented in figure \ref{fig5}(a). {The decay time was estimated from the moment of the reduction in $Re$ $\left(t=0\right)$ to the time where no disordered motion is observed within the translating disordered patch $\left(t=t_R\right)$.  In the example presented in figure \ref{fig5}(a), $t=0$ and $t=t_R$ are represented as continuous and dashed lines, respectively. The oblique streaks are related to the translation velocity along the pipe axis. The time for the disordered patch to decay, $t_R$, was extracted from the diagrams and are shown in figure \ref{fig5}(b). The straight line is a linear fit of $t_R$ indicating that the lifetime of the turbulent patch increases linearly with $Re$. A divergence of timescales is expected close to transition points\cite{peixinho06}. Our experimental setup is limited to a moderate range of $t_R$. The data suggest that a critical point for sustained localized turbulence may be found at $Re$ between 600 and 800 and is reported in figure \ref{fig2}(c).

This work presented a study about the flow in slowly diverging pipe. At low flow rate, no recirculation bubble is observed. For larger flow rate, stable laminar recirculation bubble is  observed and extends downstream. The results of our computation  predict the onset of the recirculation and the extent of the recirculation bubble for a range of parameters. 

With further increase of the flow rate, a domain of unstable turbulent patches are uncovered. The extent of the turbulent patch is reported. A future direction of our research will be to investigate the statistical properties of this localized turbulence and the puff-slug transition as the turbulent puff here does not travel along the pipe.\\

The financial support of the Japan Society for the Promotion of Science and the R\'egion Haute-Normandie are acknowledged. We also thank A. P. Willis, I. Mutabazi and J. E. Wesfreid for discussions.

\end{document}